\UseRawInputEncoding
\documentclass[pra,12pt,tightenlines,nofootinbib,floatfix]{revtex4} 
\pdfoutput=1
\usepackage{amsmath}
\usepackage{graphics, graphicx}
\usepackage{epsfig}
\usepackage{ulem}
\usepackage{color}
\usepackage{datetime} 
\begin{document}

%This has to be commented out if you want the revtex eqn style.
%\renewcommand{\theequation}{\arabic{equation}}
%One of these must be commented out
%\newcommand{\ttle}[1]{{\it #1}}
%\newcommand{\ttle}[1]{

%\begin{document}

\begin{abstract}
%\hskip .5in. 
 \centerline{ Brief recollections by the author about  how his interactions with John Wheeler influenced his career in physics.}
\end{abstract}

\title{Recollections of John Wheeler}

\author{James  Hartle}
\affiliation{Department of Physics, University of California, Santa Barbara, California  93106, USA} {\affiliation{ Santa Fe Institute, \\ 1399 Hyde Park Road,  Santa Fe, New Mexico  87501, USA.}  

\bibliographystyle{unsrt}
\bibliography{references}

%\begin{the bibliography}{99}

\maketitle

\section{Introduction}
\label{intro}

I never wrote a scientific paper with Wheeler  or worked with him extensively in science.  Despite this, my own career  was significantly influenced  by his guidance,  by his writings and  by  interactions and discussions with him. This paper reports my memories of these interactions and influences. I am relying on my memory  for these  recollections knowing  all the risks that go along with that. I  cannot record what I have forgotten in the sixty years since the story began;  I am not writing accurate history that can be backed up by documents.  I am writing  about my recollections of situations and events that occurred long ago.

As it happened  Johnny Wheeler is my uncle by marriage. I therefore have two sets of Wheeler recollections.  There is one connected with my family, and another connected with his impact on my science.  This piece is only about the second set. For more information about the first see Johnny's autobiography with Ken Ford (\it Geons, Black Holes, and Quantum Foam}, W.W.Norton, 1998). 

This is not a scientific biography of Wheeler.  For that see his US  National Academy of Sciences biographical memoir written by Kip Thorne (arXiv/1901.46623). Neither is it a personal biography of the man and the important role he played in our national life, for example his role in developing the H-bomb.  (For that see his autobiography, reference above.)   Nor does it contain any of the anecdotes about Johnny such as the one about his bomb shelter and his neighbor.  It is limited to how he influenced me.

\section{ Beginning Physics at Princeton } 
\label{princeton}  
Starting out at Princeton in  engineering school I was required  to take a first course in physics.
John Wheeler was the lecturer.  I don't remember anything especially inspiring about these lectures. They were on the basics, not on the exciting developments I read about in places like the Scientific American. But this connection  did mean that Wheeler was the person I turned to for advice in my early career and for suggestions of important problems I might work on and issues I might tackle. In today's parlance Johnny became one of my mentors.

\section{Switching to Physics} After a year in engineering I realized that it was physics that I was interested in and wanted to switch schools (engineering to liberal arts) and major in physics.  I was naively worried about as to whether I could major in physics without having taken honors freshman physics.  I sought advice from Johnny which of course was `no problem'. 

\section{Learning Mechanics on European Rail Lines}
\label{europe}
I was aware that  typical graduate schools required proficiency in two foreign languages. I thought I might get by in French but anything else was {\it really}  a foreign language. Johnny Wheeler knew of a summer school in Salzburg, Austria to learn German.  In the summer of 1958 (age 19)  I went off to Salzburg  supported by my parents.

I was an ambitious student and sought to move ahead quickly in the undergraduate physics curriculum replacing undergraduate courses with graduate ones. So for my railway journeys  I took along the undergraduate text in mechanics by Keith Symon and worked out the problems on a small errata sheet that had been left inside.  I still have my copy.

I didn't learn much German. But more importantly I learned how to travel on my own far away from home, by myself, and deal with people and situations that were new to me.   It was an adventure --- a coming of age experience. (In the end I satisfied the language requirements for a Ph.D. degree at CalTech in French and Russian.)

%In % was an ambitious student and sought to move ahead quickly. So for my jrailway journeys  I took along the undergraduate text in mechanics by Kieth Symons and worked out the problems on an errata.  I still have my copy..

Probably there was no direct connection but I'd like to think this same sprit of adventure shows up  in other decisions I made.  e.g.  in switching my main focus from engineering to physics, in effectively switching from elementary particle physics to general relativistic astrophysics,  in going to Cambridge and working with Stephen Hawking, and in  taking seriously the idea that quantum mechanics applies to the Universe as a whole.

\section{ A senior Thesis}
\label{seniorth}  
To graduate from Princeton I had to write a senior thesis --- a small piece of original research. 
%555I needed a faculty supervisor for this and approached John Wheeler.  
%He was too busy but delegated the job and the topic to his then post doc Dieter Brill. The topic was the gravitational geon and in particular the  propagation of linearized gravitational waves through a spacetime curved self-consistently by the energy of these waves.  
I sought advice from Johnny , on what to write and who to work with. (He was too busy himself.) He connected me with his postdoc Dieter Brill who later became a professor at the University of Maryland. It was a fortunate choice. Dieter had the time and patience to guide me through my first effort in independent work. Johnny also suggested the thesis topic ---  weak gravitational waves in curved background spacetimes,specifically the gravitational geon. The work is published  as (D.R. Brill and J.B.Hartle, {\it The Method of the Self-Consistent Field in General Relativity and its Application to the Gravitational Geon}, {\sl Phys. Rev. B } ,{\bf135,} 271, (1964)). It was one of my first published papers and still occasionally referred to in papers on gravitational waves.  It has become standard and the subject of ÒBrill-Hartle averagingÓ can be found described in text-books like that of Misner, Thorne, and Wheeler.

 %None of this was quantum mechanical but two aspects of it were important for my later quantum mechanical work. First, I learned general relativity which is necessary for quantum cosmology. Second, it reinforced the lesson that to achieve deep understanding its best to work things out for oneself.

\section{Graduate School} I asked advice on what graduate school to go to  from a number members of the physics faculty. Of course I asked John Wheeler.  He responded by saying that people like me who are 80 percent good should  go into solid state physics and that the University of Illinois was the best place to study that. I didn't follow that advice on either subject or place and wound up at CalTech.  

But I discovered that CalTech was not called the Marine Corps of physics for nothing and wrote him asking to return to Princeton as a graduate student.  Fig 1  contains   a modestly trimmed scan  of his supportive and kind letter in reply advising me on how to soldier on. Taking his advice I stayed at CalTech and completed my dissertation.

 \begin{figure}[t]
\includegraphics[width=6.5in]{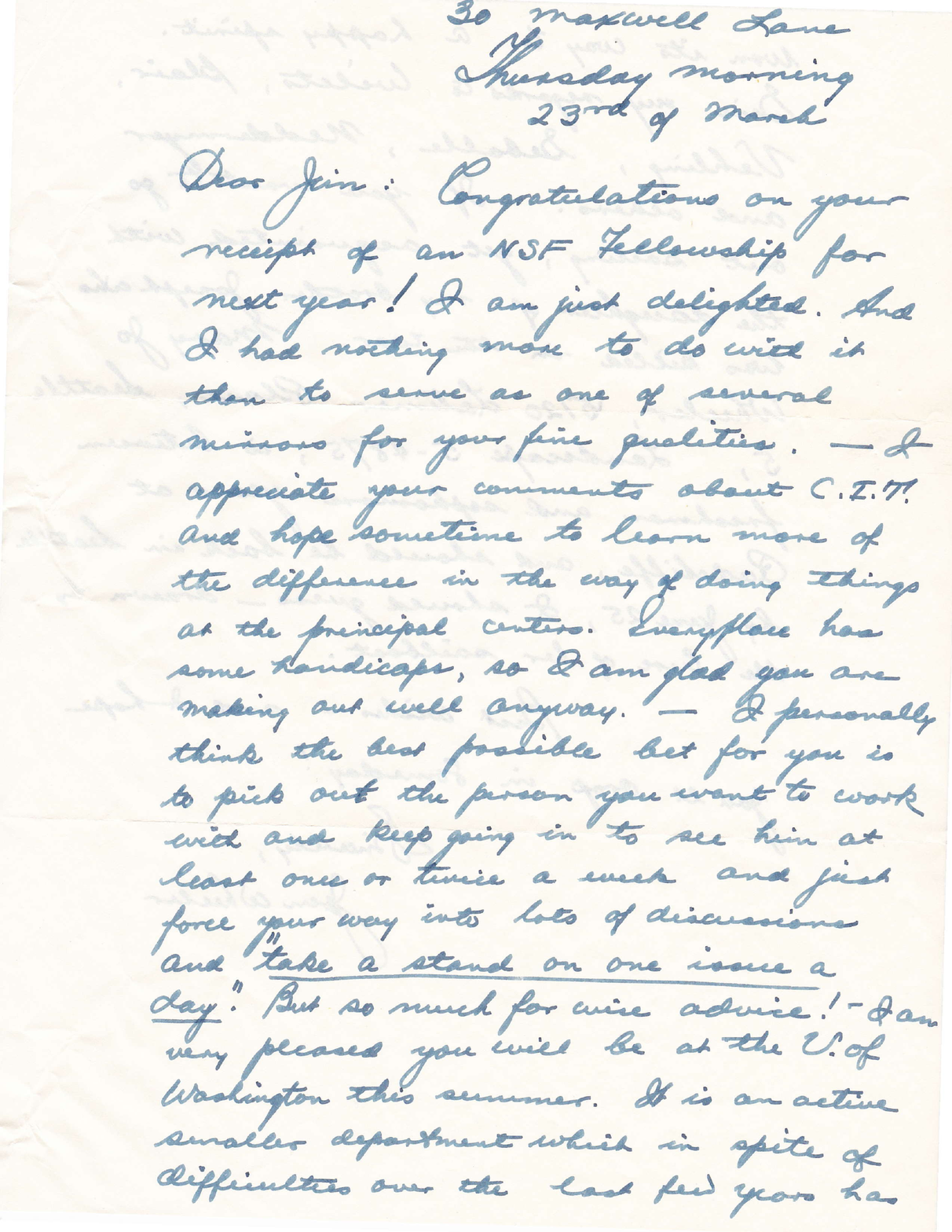}  
\caption{A scan of the main part of a letter from Johnny to me when I was a graduate student advising me how to approach potential thesis advisors.} 
\label{wheeler-letter}
\end{figure}

\section{Back in Princeton} 
\label{bkinPrinceton}
I completed my PhD with Murray Gell-Mann at CalTech in three years. In retrospect this was too fast. I had no clear idea of where to go and what to do next. When Murray asked me where I wanted to go I could only think  of the Institute for Advanced Study  (IAS) in Princeton.  A short time after  this interview the mail brought an application form to be a postdoc at the IAS. Someone had crossed out the section for my references and typed in ``We have already heard  from Professor Gell-Mann.''  I thought this looked promising. Thus I wound up back in Princeton for three years close by Johnny Wheeler.  I left only when I became an assistant professor at the University of California, Santa Barbara. 

\section{Johnny and Janette} 
\label{social} 
Dinners and events at Johnny and Janette's house served to build a group in physics and introduce younger scientists like me to a much wider range of colleagues. Janette would  ask her guests to sign their placemats.
She would then embroider them.  Mary Jo still has the place mats ---  a little bit of physics history.

%\section{General Relativistic Astrophysics} 
%\label{grastro}  Johnny urged Dieter Brill  and me to write up our work on the gravitational geon which was the subject of my undergraduate senior thesis at Princeton and a precursor to my later work in general relativistic astrophysics.  We did.   {\it Method of the Self-Consistent Field in General Relativity and its Application to the Gravitational Geon}  (with D. R. Brill), Phys. Rev. B , 135, 271, (1964). It was one of my first published papers and still occasionally referred to.

\section{Power from  Gravitational Collapse} 
\label{gravcollapse} 
Already in my beginning physics course (Physics 103),  Johnny stressed that power could be produced by gravitational collapse and even  had a lecture demonstration to support that.

The early'60's saw the discoveries of pulsars (1967), quasars (1963), and the cosmic background radiation ,(1965) --- all phenomena in which significant power is released by gravitational means and which therefore were situations where general relativity could be important. I think that Johnny saw more clearly than many that gravitational collapse would be important for producing power in the Universe and that general  relativity would be  thereby be essential for astrophysics  and advised me accordingly. 

Specifically he  encouraged  me to look into the end states of stellar evolution especially neutron stars and black holes. My papers  on variational principle for general relativistic stellar structure and stellar rotation  stemmed in part from his encouragement. This was the start of my work in general relativistic astrophysics for which I am probably best known.

\section{ A Meeting on Nuclear Physics}
\label{meeting} 
Years later,  when I was in residence at the Santa Fe Institute,  a nearby  meeting on the history of nuclear physics was announced, possibly in Albuquerque,  with Johnny and Marcos Moshinsky  as the principle speakers. I drove down down to hear Johnny  and greet him. Towards the end of his talk Johnny  broke down  in tears about the death of his brother and his talk had to be stopped,  It didn't matter. Roughly 80 people rushed forward to personally convey their thanks and best wishes, such was the affection in which he was held by many.

%\section{High Island} 
%\label{highis}
 %My wife, Mary Jo, is a niece of John Wheeler. I saw a lot of Johnny while visiting the Wheeler summer home on High Island near South Bristol, Maine. But there I never sought to argue with so great a figure who had supported me, and done so much for me, over a long period starting with his freshman physics lectures at Princeton.
 
% \section{ADM}
% \label{adm} 
%Richard Arnowitt, Stanley Deser, and Charles Misner,{\it The Dynamics of General Relativity}, arXiv:gr-qc/0405109.

\section{Quantum Theory}
\label{qmech}  Johnny gave serious thought to quantum mechanics as evidenced by his ideas captured in what he called ``It from Bit'' and the ``Participatory universe''  see e.g. (J.A. Wheeler, {\it Information, Physics , the Quantum: The Search for Links} in Proceedings  Int. Symposium: Foundations of Quantum Mechanics, Tokyo,1989.) His view was ``no phenomenon is a real phenomenon until it is an observed phenomenon''.\hskip .2in

His attitude to the decoherent histories formulation of quantum mechanics (DH) for cosmology that   that I was working on was  summed up for me in one of the many articles he wrote. I donÕt recall much about it except that is printed in a two column format. Browsing down one column I saw that it was an accurate summary of DH. He had noticed it! But the last paragraph begins with his statement that this is the kind formulation of quantum mechanics we donÕt want.
In fact,,we were miles apart. In Johnny's later vision observers produce the Universe. In DH the Universe produces observers.

Measurement situations occur throughout the Universe without the intervention of anything as sophisticated as an observer. The production of fission tracks in mica deep in the earth by the decay of a uranium nucleus leads to a measurement situation in  in which the track directions  are a record of the decay whether or not they ever registered by an observer. Such events would seem to have little effect on the creation of the Universe.

\section{A Meeting in Chicago}
\label{chicago}
 I was a professor at the University of Chicago for three three years living in Hyde Park.  By chance I sat next to Johnny  at a meeting at the Fermi Institute. I asked him``what ever happened to that attractive niece of yours?''  He replied, ``Oh she lives not  more than three blocks from here''.
The rest is history. I owe him my wife as well as pointers in physics.

\section{High Island} 
\label{highisl} As friends and relatives of John and Janette  we were invited to spend time in their summer home on High Island Maine. Johnny had a large study overlooking the ocean and worked in it every morning.  Being together in one place provided  an opportunity for me understand his views better and for him to understand mine. We discussed these in many walks around the island  but somehow never came to a unified picture.  We were too far apart. Johnny with the role of observers.  Me with  the quantum mechanics of the possible histories our Universe.

%\section{A Meeting in Chicago}
%\label{chicago} I was a professor at the University of Chicago for three three years living in Hyde Park.  I sat next to him at a meeting at the Fermi Institute. I asked him what ever happened to that attractive niece of your?  He said, ``Oh she lives not  more than three blocks from here''.
%The rest his history. I owe him my wife as well as pointers in physics.

\section{Last Visit}
\label{last}
Much later my wife Mary Jo and I were back in Princeton and able to visit John and Janette in their retirement home outside of the town. When we first greeted him he asked ``Do you have a car?'' We answered yes and he immediately replied. ``Let's get of here!''

Later  I asked him what was the greatest problem in physics that he couldn't solve? He answered immediately; ``How come existence?''. He died roughly 10 days later. I think we do not  yet have an answer to his question. And certainly  we have no one who could replace him.

%\section{A Meeting in Chicago}
%\label{chicago} I was a professor at the University of Chicago for three three years living in Hyde Park.  I sat next to him at a meeting at the Fermi Institute. I asked him what ever happened to that attractive niece of yours?  He responded , ``Oh she lives not  more than three blocks from here''.
%The rest is history. I owe him my wife as well as pointers in physics.

\section{Conclusion}
\label{conclu} 
Through his vision, guidance and support John Wheeler played an important role in my career, 
as he did with many others.  Even today his vision is sometimes in  my mind. But however often that  happens  he will also be --- always  in my heart. 

\section{Acknowledgements} 
\vskip .1in 
\noindent{\bf Acknowledgments:}    Thanks  are due to the NSF for supporting its preparation under grant PHY-18-8018105 and to Mary Jo Hartle for proofreading it more than once.

 \end{document}